\documentstyle[12pt]{article}
\begin{document}
\baselineskip4mm
\title{\vspace{-3cm} 
{\small \hfill{
\begin{tabular}{l}
DSF$-$97/20 \\
gr-qc/9705043\\
$~$
\end{tabular}}}\\
Quantum Cosmology and Grand Unification}
\author{G. Esposito$^{1,2}$, \ A. Yu. Kamenshchik$^{3}$ and
\ G. Miele$^{2,1}$}
\date{}
\maketitle
\hspace{-6mm}$^{1}${\em Istituto Nazionale di Fisica Nucleare,
Sezione di Napoli, Mostra d'Oltremare Padiglione 20, 
80125 Napoli, Italy}\\
$^{2}${\em Dipartimento di Scienze Fisiche, Mostra
d'Oltremare Padiglione 19, 80125 Napoli, Italy}\\
$^{3}${\em Nuclear Safety Institute, Russian
Academy of Sciences, 52 Bolshaya Tulskaya, Moscow 113191, Russia}\\

\begin{abstract} Quantum cosmology may restrict the class of
gauge models which unify electroweak and strong interactions.
In particular, if one studies the normalizability criterion
for the one-loop wave function of the universe in a de Sitter
background one finds that the interaction
of inflaton and matter fields, jointly with the request
of normalizability at one-loop order, picks out
non-supersymmetric versions of unified gauge models.
\end{abstract}

The investigations in modern cosmology have been devoted to
two main issues. On one hand, there were the attempts to
build a quantum theory of the universe with a corresponding
definition and interpretation of its wave function 
[1,2]. On the
other hand, the drawbacks of the cosmological standard model
motivated the introduction of inflationary scenarios. These 
rely on the existence of one or more scalar fields, and a
natural framework for the consideration of such fields is
provided by the current unified models of fundamental
interactions (see, for example, Ref. [3] and references
therein). The unification program started with 
the proposal and the consequent experimental verification of the 
electroweak standard model ($SU(3)_C \otimes SU(2)_L \otimes U(1)_Y$),
and has been extended to other simple 
gauge groups, like $SU(5)$, $SO(10)$ 
and $E_6$. All of them in fact, even if with different capability,
unlike the electroweak standard model 
are able to allocate all matter fields in a few 
irreducible representations (IRR) of the gauge group,
and require a small number of free parameters. However, since these
enlarged gauge models predict new physics, a first source of constraints
upon them is certainly provided by the experimental bounds on processes like
proton decay, neutrino oscillations, etc. [4]. Further restrictions
can be obtained from their cosmological applications, as
discussed in Ref. [5]. 

One can say, however, that the majority of investigations,
studying the mutual relations between particle physics and
cosmology, leave quantum cosmology itself a bit aside, using
it only as a tool to provide initial conditions for inflation.
Meanwhile, one can get some important restrictions on particle
physics models, using the general principles of quantum theory.
In particular, we study the possible
restrictions on unified gauge models resulting from a
one-loop analysis of the wave function of the universe and
from the request of its normalizability [6--10]. It is known that
the Hartle-Hawking wave function of the universe [1], as 
well as the tunnelling one [2], are not normalizable at
tree level [11]. In Ref. [6] it was shown that, by taking 
into account the one-loop correction to the wave function,
jointly with a perturbative analysis of cosmological 
perturbations at the classical level, one can obtain a
normalizable wave function of the universe provided that
a restriction on the particle content of the model is fulfilled.
Such a restriction is derived from the formula for the
probability distribution for values of the inflaton
field [6]
\begin{equation}
\rho_{HH,T}(\varphi) \cong {1\over H^{2}(\varphi)}
{\mbox e}^{\mp I(\varphi)
-\Gamma_{1-\rm{loop}}(\varphi)} \; ,
\end{equation}
where the subscripts 
HH and T denote the Hartle-Hawking and tunnelling wave
function, respectively, $H(\varphi)$ is the effective Hubble
parameter, $\Gamma_{1-\rm{loop}}$ is the one-loop effective
action on the compact de Sitter instanton. One can show from (1),
that the normalizability condition of the probability 
distribution at large values of the inflaton scalar field
$\varphi$ is reduced to the condition [6]
\begin{equation}
Z > -1 \; ,
\end{equation}
where $Z$ is the total anomalous scaling of the theory. This
parameter is determined by the total Schwinger-DeWitt
coefficient $A_{2}$ in the heat-kernel asymptotics [12],
and depends on the particle content.

In Ref. [7] the criterion (2) was used to investigate the
permissible content of different models. It was noticed
that the standard model of particle physics, as well as
the minimal $SU(5)$ GUT model, does not satisfy the criterion
of normalizability, while the standard supersymmetric model,
the $SU(5)$ SUSY model and $SU(5)$ supergravity model do
satisfy this criterion. 

All the analysis in Ref. [7] was carried out in terms of
physical degrees of freedom, e.g. three-dimensional transverse
photons or three-dimensional transverse-traceless metric
perturbations. However, over the last few years, the explicit
calculations have shown that a covariant path integral for
gauge fields and gravitation yields an anomalous scaling which
differs from the one obtained from reduction to physical
degrees of freedom. This holds both for closed manifolds and
for manifolds with boundary [12--14]. 

Unfortunately, the reduction to physical degrees of freedom
is not well defined on any curved Riemannian manifold [12].
Moreover, such a reduction does not take explicitly into
account gauge and ghost terms in the path integral, and leads
to a heat-kernel asymptotics which disagrees with the
well-known results of invariance theory. For all these
reasons, we regard the covariant version of the path integral
as more appropriate for one-loop calculations.

In Ref. [8] the investigation of the one-loop wave function
was carried out for a non-minimally coupled inflaton field
with large negative constant $\xi$. It was then shown that 
the behaviour of the total anomalous scaling $Z$ is determined
by interactions between the inflaton and remaining matter
fields. 

Here, we study normalizability properties of a wide set of
unified gauge models, with or without interaction with the
inflaton field. The models studied are, as shown in Table I,
the standard model of particle physics, $SU(5)$, $SO(10)$
model in the 210-dimensional irreducible representation,
$E_{6}$, jointly with supersymmetric versions of all these
models with or without supergravity. The building blocks of
our one-loop analysis are the evaluations of $A_{2}$ coefficients
for scalar, spinor, gauge, graviton and gravitino perturbations.
All these coefficients (but one) are by now well-known,
and are given by
\begin{eqnarray}
A_{2 \; \rm {scalar}}&=&
{29\over 90}-4\xi+12 \xi^{2} \nonumber\\
&-& {1\over 3}m^{2}R_{0}^{2}
+2\xi m^{2} R_{0}^{2}
+{1\over 12}m^{4}R_{0}^{4} \; ,
\end{eqnarray}
\begin{equation}
A_{2 \; {\rm {spin}}-1/2}={11\over 180}+{1\over 3}
m^{2}R_{0}^{2}+{1\over 6}m^{4}R_{0}^{4} \; ,
\end{equation}
\begin{equation}
A_{2 \; \rm {gauge}}=-{31\over 45}+{2\over 3}m^{2}
R_{0}^{2}+{1\over 3}m^{4}R_{0}^{4} \; ,
\end{equation}
\begin{equation}
A_{2 \; \rm {gravitino}}=-{589\over 180} \; .
\end{equation}
It should be stressed that Eq. (3) only holds for scalar
fields different from the inflaton. With our notation,
$m,\xi$ and $R_{0}$ represent effective mass, (dimensionless)
coupling parameter, and 4-sphere radius, respectively.
Equation (4) holds for a spin-1/2 field with half the number
of modes of a Dirac field. Since the results (5) and (6) rely
on the Schwinger-DeWitt technique, they incorporate, by
construction, the effect of ghost zero-modes. However, it has
been argued in Ref. [15] that zero-modes should be excluded 
to obtain an infrared finite effective action which is smooth
as a function of the de Sitter radius on spherically symmetric
backgrounds. On the other hand, the prescription which includes
ghost zero-modes makes the one-loop results continuous. Strictly,
we are considering small perturbations of a de Sitter background
already at a classical level (see [6--10]). There are also deep
mathematical reasons for including zero-modes, and they result
from the spectral theory of elliptic operators. Thus, we
use the expressions (5) and (6).

Last, the contribution of gravitons to the total $Z$ should be
calculated jointly with the inflaton contribution. What happens
is that the second-order differential operator given by the
second variation of the action with respect to inflaton and metric
is non-diagonal even on-shell, by virtue of a non-vanishing
vacuum average value of the inflaton [16,17]. The resulting
$A_{2}$ coefficient turns out to be independent of the value
of $\xi$ and equal to [10]
\begin{equation}
A_{2 \; {\rm {graviton}} + {\rm {inflaton}}}=-{171\over 10} \; .
\end{equation}
In Table I, we report the total $Z$ for some relevant
examples of GUT theories, whenever one neglects the 
mass terms. This ansatz is correct, if the interaction
between inflaton and the other particles is not considered. In this
case in fact, the term $m^{2}
R_{0}^{2} \sim\varphi^{-2}$ is very small by virtue of
the large value of $\varphi$. The analysis starts with the 
electroweak standard model (SM), which contains, in its non-SUSY
version, $45$ Weyl spinors (we neglect for simplicity right-handed
neutrinos and their antiparticles), $24$ gauge bosons and one doublet 
of complex Higgs fields. The particle content changes for the SUSY
version of this model in its minimal form (MSSM) [18]. 
In this case, in fact, 
to the $45$ Weyl leptons and quarks one has to add $4$ higgsinos 
and $12$ gauginos, whereas the scalar sector consists now of 
$90$ sleptons and squarks plus $8$ real scalar fields. 
A similar analysis is performed for the
$SU(5)$ GUT model [19],
which in its non-SUSY version, apart from the $24$
gauge bosons, needs scalars belonging to $\underline{24} \oplus 
\underline{5} \oplus \underline{\overline{5}}$ IRR's to accomplish 
the spontaneous symmetry breaking pattern. 
The matter content of the SUSY extension of the 
model [20] is obtained by 
doubling the number of Higgs IRR's used, and by adding superpartners
to any degrees of freedom. 
As far as $SO(10)$ gauge theories are concerned, 
we have considered the particular model containing 
$\underline{210} \oplus (\underline{126} \oplus \underline{\overline{126}})
\oplus \underline{10} \oplus \underline{10}$ IRR's of Higgs fields,
which is still compatible with the present experimental limit on the 
proton lifetime and neutrino phenomenology [4]. Furthermore, we have 
also considered the SUSY extension of $SO(10)$, which, to be 
consistent also with cosmological constraints, needs complex   
Higgs fields belonging to $\underline{1} \oplus \underline{10}
\oplus \underline{10}' \oplus \underline{45} 
\oplus \underline{45}' \oplus \underline{54} \oplus
\underline{54}' \oplus \underline{126} \oplus \underline{126}'$
IRR's [21].
Last, we have also considered $E_6$ GUT theories, for which
fermions are allocated in three $\underline{27}$ fundamental IRR's,
and scalars belong to two $(\underline{78} \oplus \underline{27}   
\oplus \underline{351})$ [22]. For the SUSY extension of
this model, we have just added the superpartner degrees of freedom.
Concerning the SUGRA versions of all the above models, they
have been obtained from the supersymmetric ones, just by adding 
the gravitino contribution (i.e. subtracting the $A_{2}$
coefficient in Eq. (6), because of the fermionic statistics).
Indeed, we have considered particular versions of $SO(10)$ and
$E_{6}$ gauge models, but we expect that the qualitative 
features of the results should remain unaffected.

In Table I, we have assumed that one of the Higgs fields plays
the role of the inflaton. 
The forbidden range denotes the range of values
of $\xi$ for which the normalizability criterion (2) is not
satisfied. Interestingly, conformal coupling 
(i.e. $\xi=1/6$) is ruled out
by all 12 models listed in Table I. Moreover, for the
standard and $SU(5)$ models, minimal coupling (i.e.
$\xi=0$) is also ruled out. At this stage, supersymmetric
models are hence favoured, as well as non-supersymmetric
models with a large number of scalar fields.

In the formulation of physical models, however, one has to
move gradually from the original, simplified case, towards
a more involved problem which is physically more realistic.
In our investigation, this means having to deal with the
interactions between the inflaton and remaining fields,
since such interactions are responsible for the reheating
in the early universe. This is a stage as important as the
inflationary phase. Indeed, as shown in Refs. [8,10], for a
scalar field with mass $m_{\chi}$ and constant $\xi_{\chi}$
of non-minimal interaction (which differs from $\xi$ in
Eq. (3)), one finds on a de Sitter background
\begin{equation}
\zeta_{\chi}(0)={29\over 90}-4\xi_{\chi}+12\xi_{\chi}^{2}
-{1\over 3}{m_{\chi}^{2}\over H^{2}}
+{1\over 12}{m_{\chi}^{4}\over H^{4}},
\end{equation}
where $m_{\chi}^{2}={\lambda_{\chi}\varphi_{0}^{2}\over 2}$.
Moreover, for a spin-1 gauge field with mass $m_{A}$, and a
massive Dirac field with mass $m_{\psi}$, one finds [8,10]
\begin{equation}
\zeta_{A}(0)=48 \xi^{2}{g_{A}^{2}\over \lambda^{2}}
\left[1+{(1+2\delta)\over 4\pi}
{m_{p}^{2}\over {\mid \xi \mid} \varphi_{0}^{2}}
+O(1/ \mid \xi \mid) \right],
\end{equation}
\begin{equation}
\zeta_{\psi}(0)=-48\xi^{2}{f_{\psi}^{2}\over \lambda^{2}}
\left[1+{(1+2\delta)\over 4\pi}
{m_{p}^{2}\over {\mid \xi \mid} \varphi_{0}^{2}}
+O(1/ \mid \xi \mid) \right],
\end{equation}
where the coupling constants $g_{A}$ and $f_{\psi}$ are
related to the masses by the formulas
$m_{A}^{2}=g_{A}^{2}\varphi_{0}^{2}$,
$m_{\psi}^{2}=f_{\psi}^{2}\varphi_{0}^{2}$, and the
parameter $\delta$ is defined by 
$ \delta \equiv -{8\pi \mid \xi \mid m^{2}\over 
\lambda m_{p}^{2}}$, $\lambda$ being the parameter of
self-interaction for the inflaton. Thus, if one considers 
supersymmetry, jointly with a Wess-Zumino scalar multiplet
interacting with the inflaton, the terms of order
$m^{4}R_{0}^{4}$ in Eqs. (3) and (4) cancel each other 
exactly after combining contributions proportional to [10]
$$
\sum_{\chi}\lambda_{\chi}^{2}
+16 \sum_{A}g_{A}^{4}
-16 \sum_{\psi} f_{\psi}^{4}.
$$
By contrast, terms of order $m^{2}R_{0}^{2}$ have opposite
signs, since they are proportional to
$$
-8 \sum_{\chi}\lambda_{\chi}
+32 \sum_{A}g_{A}^{2}
-32 \sum_{\psi} f_{\psi}^{2}.
$$
At this stage, one has to bear in mind that, by virtue of
cosmological perturbations, one can prove that $m^{2}R_{0}^{2}$
is of order $10^{4}$ [23]. The effect of all these properties
is hence a negative value of $Z$ which cannot be greater than
$-1$ (cf. Eq. (2)). In other words, inflaton interactions
revert completely the conclusions that, otherwise, would be
drawn from Tab. I. In particular, our analysis proves that
the ``pseudo-supersymmetric" combination of coupling constants
considered in Refs. [8--10] does not improve the situation
with respect to the criterion in Eq. (2).

Our investigation shows that the one-loop normalizability
criterion for the wave function of the universe picks out
non-supersymmetric versions of unified gauge models [24].
Despite this negative result, the investigation of 
supersymmetric cosmological models remains an important
task, at least from the point of view of the general
formalism of modern field theories [25,26]. Moreover, the
problem remains of proving that our conclusion is not
affected by higher-order effects in the semiclassical
evaluation of the wave function of the universe. As far
as we know, these effects cannot be studied with the help
of $\zeta$-function methods, and represent a fascinating
problem in the quantum theory of the early universe.

A. K. is indebted to A. Barvinsky for collaboration and
useful discussions on these topics.
A. K. was partially supported by RFBR via grant no. 
96-02-16220 and RFBR-INTAS via grant no. 644, and by Russian
research project ``Cosmomicrophysics".

\footnotesize
\bigskip\bigskip
\par\noindent
{\bf Table I.}\\
\bigskip
\begin{tabular}{|c|c|c|c|}
\hline
 & & & \\
Gauge group & version & $Z$ & forbidden $\xi$ range \\
 & & & \\  
\hline   
& & & \\
& non-SUSY & $36 \xi^2 -12 \xi - {543 \over 20}$ & 
$-.701 \leq \xi \leq 1.035$ \\
& & & \\  
$SU(3)_C \otimes SU(2)_L \otimes U(1)_Y$ & SUSY & 
$1164 \xi^2 -388 \xi + {389 \over 180}$ &  
$.008 \leq \xi \leq .325$ \\
& & & \\  
& SUGRA & $1164 \xi^2 -388 \xi +{163 \over 30}$ & 
$.017 \leq \xi \leq .316$ \\   
& & & \\  
\hline 
& & & \\  
& non-SUSY & $ 396 \xi^2 - 132 \xi -{103 \over 4} $ & 
$-.134 \leq \xi \leq .467 $ \\
& & & \\  
$SU(5)$ & SUSY & 
$1884 \xi^2 - 628 \xi +{1919 \over 180} $ &  
$.020 \leq \xi \leq .314$ \\
& & & \\  
& SUGRA & $ 1884 \xi^2 -628 \xi +{209 \over 15}$ & 
$ .026 \leq \xi \leq .308$ \\   
& & & \\
\hline 
& & & \\
& non-SUSY & $ 5772 \xi^2 - 1924 \xi + {4678 \over 45} $ & 
$.069 \leq \xi \leq .265 $ \\
& & & \\  
$SO(10)$ & SUSY & 
$ 12444 \xi^2 - 4148 \xi + {11321 \over 45} $ &  
$.080 \leq \xi \leq .253$ \\
& & & \\  
& SUGRA & $ 12444 \xi^2 - 4148 \xi +{5097 \over 20}$ & 
$.082 \leq \xi \leq .252$ \\   
& & & \\  
\hline 
& & & \\  
& non-SUSY & $ 10932 \xi^2 -3644 \xi +{39197 \over 180} $ & 
$.078 \leq \xi \leq .255 $ \\
& & & \\  
$E_6$ & SUSY & 
$ 12876 \xi^2 - 4292 \xi +{42719 \over 180} $ &  
$.070 \leq \xi \leq .263$ \\
& & & \\  
& SUGRA & $ 12876 \xi^2 -4292 \xi +{1203 \over 5}$ & 
$.072 \leq \xi \leq .262$ \\   
& & & \\
\hline 
\end{tabular}
\normalsize   

\end{document}